# Review on the Design of Web Based SCADA Systems Based on OPC DA Protocol

**Hosny A. Abbas**                                                hosnyabbas@yahoo.com
*Senior automation Engineer*
*Qena Paper Company*
*Qus, Qena, Egypt,P.O:83621*

**Ahmed M. Mohamed**                                         ahmed@engr.uconn.edu
*Electrical Eng. Department, Aswan Faculty of*
*Engineering*, *South Valley University*
*Aswan, Egypt*

**Abstract**

One of the most familiar SCADA (supervisory control and data acquisition) application protocols now is OPC protocol. This interface is supported by almost all SCADA, visualization, and process control systems. There are many research efforts tried to design and implement an approach to access an OPC DA server through the Internet. To achieve this goal they used diverse of modern IT technologies like XML, Webservices, Java and AJAX. In this paper, we present a complete classification of the different approaches introduced in the litrature. A comparative study is also introduced. Finally we study the feasibilty of the realization of these approaches based on the real time constraints imposed by the nature of the problem.

## 1. INTRODUCTION
OPC stands for OLE (Object linking and Embedding) for Process Control - now it also stands for Open Process Control- draws a line between hardware providers and software developers. It provides a mechanism to provide data from a data source and delivers the data to any client application in a standard way. The utility of OPC has now reached the point where automation without OPC is unthinkable. This interface is supported by almost all SCADA, visualization, and process control systems [1]. It gives production and business applications across the manufacturing enterprise access to real-time plant floor information in a consistent manner, making multivendor interoperability and "plug and play" connectivity a reality [2]. Interoperability is assured through the creation and maintenance of open standards specifications. There are currently seven standards specifications completed or in development. Based on fundamental standards and technology of the general computing market, the OPC Foundation adapts and creates specifications that fill industry-specific needs. OPC will continue to create new standards as needs arise and to adapt existing standards to utilize new technology. OPC is based on the DCOM/COM component-object programming model developed by Microsoft in which software is divided into smaller, independent units (the objects). Web-based SCADA system uses the Internet to transfer data between the RTUs (Remote Terminal Units) and the MTU (Master terminal Unit) and/or between the operators' workstations and the MTU. This will reduce the cost of the installation of the SCADA network if compared with installing a dedicated network [3]. Therefore, many researchers all over the world tried to design and to implement an approach to access an OPC DA server through internet to realize a web based SCADA system. DCOM is suitable for LANs where there are less interruptions and noise, but when used through Internet there will be some limitations related to its nature. For this reason, the researchers tried to use IT (Information Technologies) services to achieve their goals. In the following sections, we present an overview of the work in this area.

## 2. Previous Work

The main challenge will be accessing an OPC DA server, which is a COM/DCOM server through the Internet. We have two options, using DCOM or using modern IT technologies. There are many solution presented in each option. We classify them into the following categories.

### 2.1 DCOM

Most of the previous research used DCOM for communication with OPC DA server through LANs. For example Xiaofeng Lee et al. [4] uses DCOM communication between an OPC DA client and OPC DA server, then they transfer the OPC DA client data to XML format to be able to access these data through Internet with an XML-DA client which communicates to an XML server (Web server) to get data as shown in Figure 1. Also Truong Chau et al. [5] uses DCOM to enable the C# server script to access OPC DA through LAN and because the OPC DA client is a .NET client, they used an OPC .NET wrapper to make the transformation from .NET to COM and COM to .NET as shown in Figure 2. Zhang Lieping et al. [6] uses the OPC DA Toolbox which is integrated in MATLAB 7 and above editions, this Toolbox enables MATLAB applications to communicate with OPC DA servers through DCOM then the user can simply and conveniently realize the operation to the OPC objects. As they claimed that these features could simplify the process of development and provide an effective method to realize the remote real-time communication between MATLAB and process devices as shown in Figure 3. From the above discussion, we conclude that the only way to communicate directly to the OPC DA server is DCOM (or COM if the client and server are on the same machine). Unfortunately using DCOM through Internet is avoided for many reasons such as, DCOM is windows dependent platform, difficult to configure, has very long and non-configurable timeouts, and cannot be used for Internet communication. That is why DCOM is best suitable for LANs where there are less number of nodes and small delay times. Therefore, the first step to access an OPC DA through web is to use COM or DCOM through LAN, and then we have to do a suitable transformation to enable accessing the OPC data through Internet.

### 2.2 XML

XML is a platform-independent, which is an important feature to achieve interoperability between different applications, which are running on different platforms. This is the reason, which forced the OPC FOUNDATION to release the OPC XML-DA specifications to allow the XML applications to access OPC data in a standard way. The other advantage of OPC XML-DA is the simple administration as it based on SOAP and XML. On the other hand, it has some disadvantages such as:
- Not suitable for transferring large data volumes
- XML technology is generally slower than COM

- The interaction parameters coded using XML, which leads to an overhead.
- An OPC XML-DA Service is stateless.

During browsing, no information about the position of the client in the namespace is stored in the OPC XML-DA Service, but all information about the namespace (or a defined part of it) transferred to the client at the same time. The client can poll for values at the server, but it should also be possible, to receive changed values automatically. XML-DA servers may stand alone, or may be developed to wrap COM based 3.0, and even 2.0x servers [7] [4].

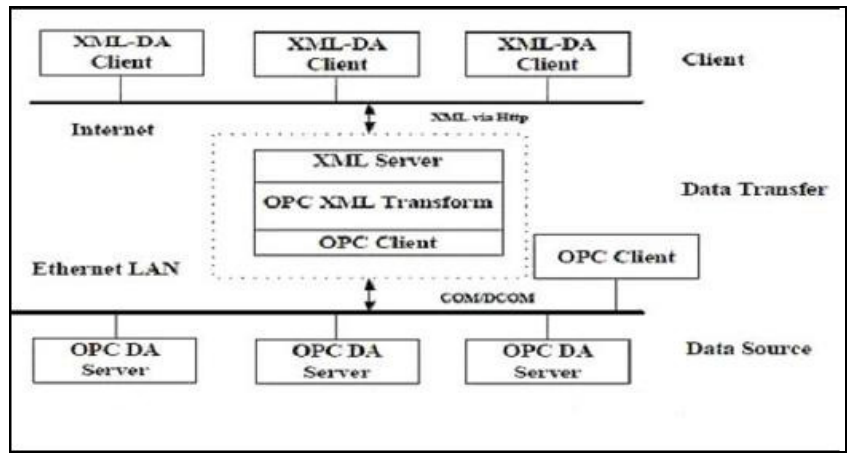

**FIGURE 1 :** Xiaofeng Lee et al. [4] Design

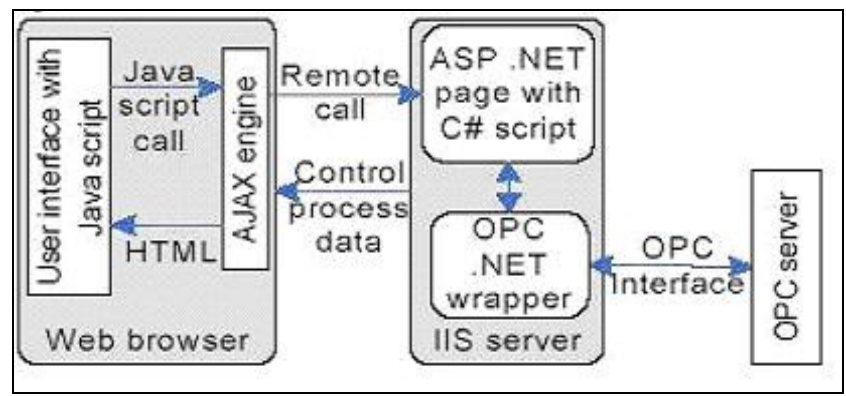

**FIGURE 2 :** Truong Chau et al. [5] Design

Xiaofeng Lee et al. [4] suggested designing an information integration system which will adopt OPC DA to OPC XML-DA; the design includes three layers structure, data source layer, data transfer layer and client layer as shown in Figure 1. The authors claim that because of adopting industry standard OPC interface and webservice transfer interface, the remote monitoring system based on OPC XML-DA technology makes it convenient to update and expand system. If we analyze this approach, we will find that there is an overhead in layer2 because of the COM-XML transformer. In addition, the authors did not expose to the problem of client data update, is there a data polling mechanism that enables the client to get the new data- if there is –in an efficient way that consumes as little as possible of available resources. For similar work, that uses XML and/or OPC XML-DA techniques see [8, 9, 10]. Due to possible performance limitations, OPC XML-DA is unlikely used for real time applications, although it is commonly used as a bridge between the enterprise and control network. Furthermore, only OPC-DA functionality is provided in XML-DA. So it can be best seen as a transitional path to a true Webservice architecture that is just released by the OPC Foundation, which is OPC-UA (Unified Architecture) project [11].

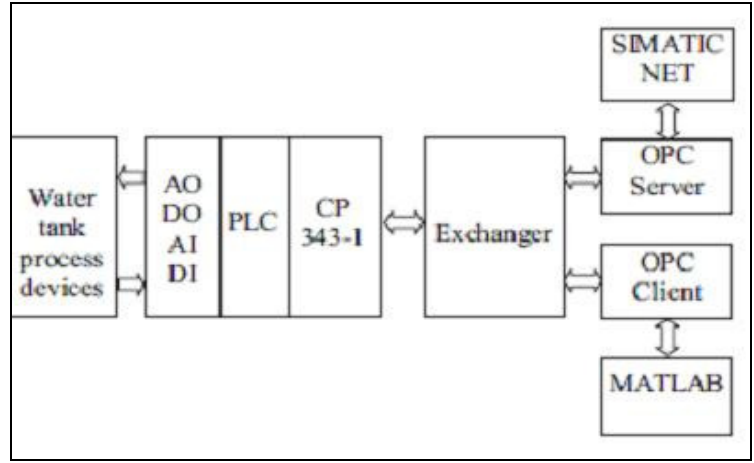

**FIGURE 3:** Zhang Lieping et al. [6] Design

**2.3 Webservices**
Webservices as defined by the W3C as "a software system designed to support interoperable machine-to-machine interaction over a network. As Shekhar M. Kelapure et al. [12] mentioned about the Webservices features, which are:
- Communicate via open protocols (HTTP, SMTP, etc.)
- Processes XML messages framed using SOAP
- Describes its messages using XML Schema
- Provides an endpoint description using WSDL
- Can be discovered using UDDI

The authors found that they could achieve a good web based SCADA system using Webservices as shown in Figure 4. They concluded that the advantages obtained when using Webservices in SCADA systems are:
- Web services provide data publishing on the internet through HTTP protocol, thereby eliminating the need to compromise on security of SCADA servers.
- Webservices provide the means to publish the data through various devices like thin clients, PDAs and mobile phones.
- Webservices facilitate interfacing of multiple control centers, like ICCP, once the services are standardized.
- Webservices are supported by .NET as well as JAVA technologies.
- Data can be fetched from Hard Disc as well as RAM using Web-services

On the other hand, Webservices have some disadvantages like:
- Webservices standards for features such as transactions are currently nonexistent or still in their infancy compared to more mature distributed computing open standards such as CORBA.

- Webservices may suffer from poor performance compared to other distributed computing approaches such as RMI, CORBA, or DCOM. This is a common trade-off when choosing text-based formats. XML explicitly does not count among its design goals either conciseness of encoding or efficiency of parsing.
- In addition, as mentioned in [13] that by utilizing HTTP, webservices can evade existing firewall security measures whose roles are intended to block or audit communication between programs on either side of the firewall.

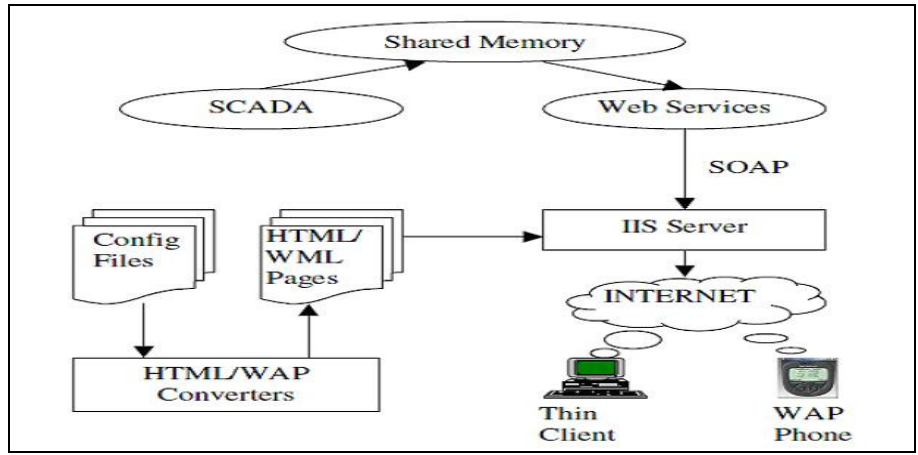

**FIGURE 4:** Shekhar M. Kelapure et al. [12] design

Nunzio M. Torrisi et al. [14] proposed what they called CyberOPC which is a communication system anticipates the use of a gateway station called CyberOPC gateway, which will process messages sent to the OPC towards the public network and vice versa. Their proposed communication system targeted to best effort network with minimum bandwidth reserved for periodic traffic as shown in Figure 5.

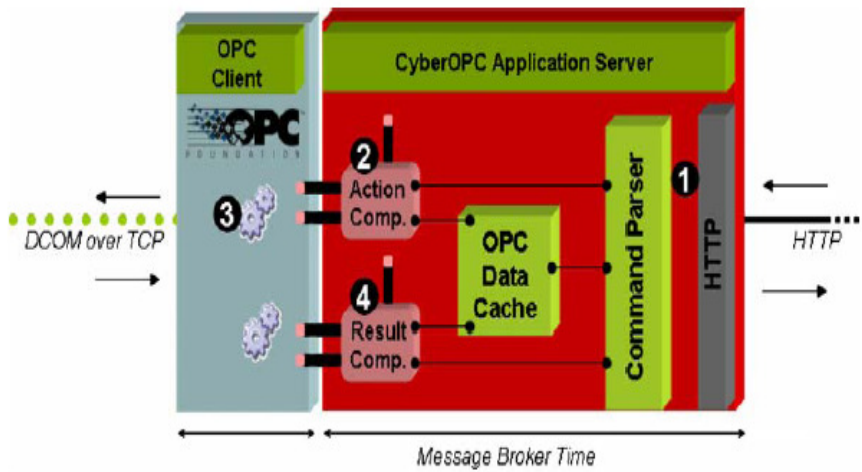

**FIGURE 5 :** Nunzio M. Torrisi et al. [14] Design

As they claimed that the necessity to satisfy time-critical and security requirements for remote control has stimulated the study of a new protocol for process control. Moreover, to obtain maximum interoperability with existing factory floor technologies, they built their communication project over the OPC technology. Unlike many OPC gateways for the Internet that use Webservices with HTTP and SOAP, the CyberOPC gateway does not use Webservices because the loss in performance will not be balanced by the advantage of a high-level programming language offered by Webservices. However, in order to use Webservices that transport OPC data, OPC libraries for processing OPC messages are required. Therefore, assuming that the use of libraries is necessary for processing non-open source and usually not free OPC data, they believe that it is useful to develop a set of free and open source libraries in order to implement the OPC communication over the Internet with the best possible performance. This approach does not solve the problem of periodic data update of the remote client to get new data. Moreover, when the remote client makes a request, the CyberOPC will not check if there is a change in this data from the last sent data or not, so if the control process has a high frequency data change rate, the remote client has to increase its periodic data-requesting rate, which can affect the server efficiency and network bandwidth.

**2.4 Classic Approaches**

Thomas Bauer et al. [15] claimed that to integrate OPC technology with Internet, its necessary to have an asynchronous data transmission method that is time-independent bidirectional communication, which is not possible because of the architecture of Internet. Therefore, they suggested an approach to solve this problem as shown in Figure 6. As they mentioned, first the Client (browser) will initiate a connection request for establishing at least one transmission channel that should be permanently open to send data at any desired time (asynchronously) and independently of the action of the user. In order to keep this data connection permanently open the web server should continuously send data to the client, if there is no useful data it will send dummy data or send information to tell the client that a useful data coming. The need of dummy data is to maintain the data connection. Keeping this channel permanently open will enable the server to initiate the communication to the client, in the same time the client is still can initiate a new request to the server using a another channel. In other words, the permanent channel will be for the server and the other channels are for the client. The main disadvantage of this approach is that the server and network will suffer because of the continuous transmit of useful and unuseful data. What will happen if the data change rate is slow? The web server will send a dummy data for long time wasting the network bandwidth with no feasibility. Also what will happen if for some reason the permanently open channel broken? The client will have to reinitiate the channel again. In addition to that, the client must keep checking the permanent open channel for new useful data the action that can affect the user interactivity.

Duo Li et al. [3] suggested an approach to implement the function of real time monitoring which provides periodically updated data to operators. The target data, which saved in a database, is marked and mapped to an HTML file-, say file2- in a web server; the database server automatically refreshes file2 with the latest data whenever the target data in the server updated. To show target data in an HTML file a Java applet developed that connects to file2 to get the latest data periodically and display them in the required format, another HTML file -say file1- in which the applet is included, developed to establish a basic human-machine interface (HMI) and complete some initiations. Data monitor functions commence from a web browser sending an HTTP GET command to the web server asking for file1, which then fetched to the browser and displayed and the applet is downloaded to run to show the target data as shown in Figure 7.

Hosny A. Abbas & Ahmed M. Mohamed

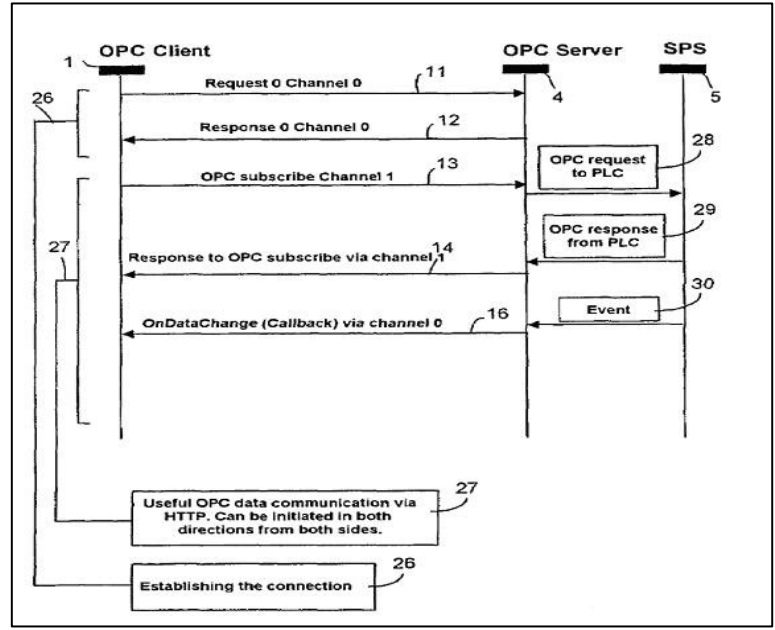

**FIGURE 6 :** Thomas Bauer et al. [15]

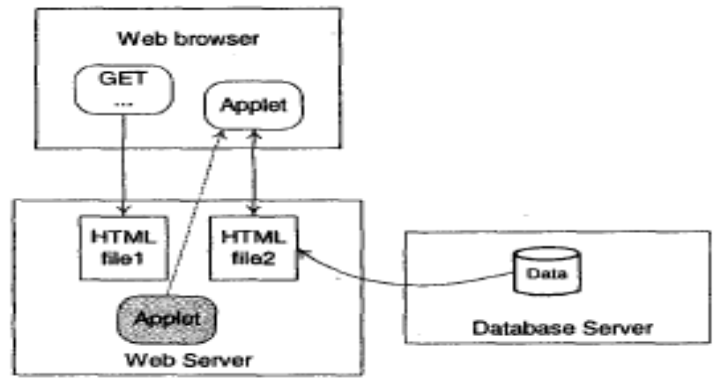

**FIGURE 7 :** Duo Li et al. [3] Approach

Actually, Java is a perfect programming language especially for web application development because its programs after compiled transferred to what known as byte code format, which can executed by Java virtual machine (JVM) that is now supported by most platforms so a good interoperability can be achieved by Java. In addition, Java has many other features like Java applets and Servlets. Applets run on client (browser) and Servlets run on a web server. Applets and Servlets can communicate with each other in an efficient and persistent way, so they used to develop web based SCADA systems. On the other hand, they have some disadvantages:
1. Java plug-in is required to run applet
2. Java applet requires JVM so first time it takes significant startup time.
3. If the applet not already cached in the machine, it downloaded from Internet and will take time.
4. Its difficult to design and build good user interface in applets compared to HTML technology

Disadvantages of Servlets
1. Developers MUST know JAVA.
2. Web Administrator will need to learn how to install and maintain Java Servlets
3. Tedious uses of out.println() statements
4. Can be remedied by using Java Server Page (JSP)

Mike Clayton et al. [16] designed a TCP client-server Java architecture based on socket support to develop applications to integrate SCADA systems and Web applications. As he mentioned, that he achieved a solution to have a bridge between two complex applications (SCADA software and web servers), which evolve independently because they belong to different worlds. The proposed design is shown in Figure 8, which illustrates the TCP socket communication between the web server and the SCADA system. As shown in the figure that there is a Java application on the SCADA system machine, this Java application is a server, which allows the TCP connections from the remote clients. The Java application or the server communicate with the SCADA system by JNI (Java native interface) which enables Java classes to communicate with standard programs using custom librarries in C/C++. For each client TCP connection request from the web server, the Java server will create a new thread to handle that connection.

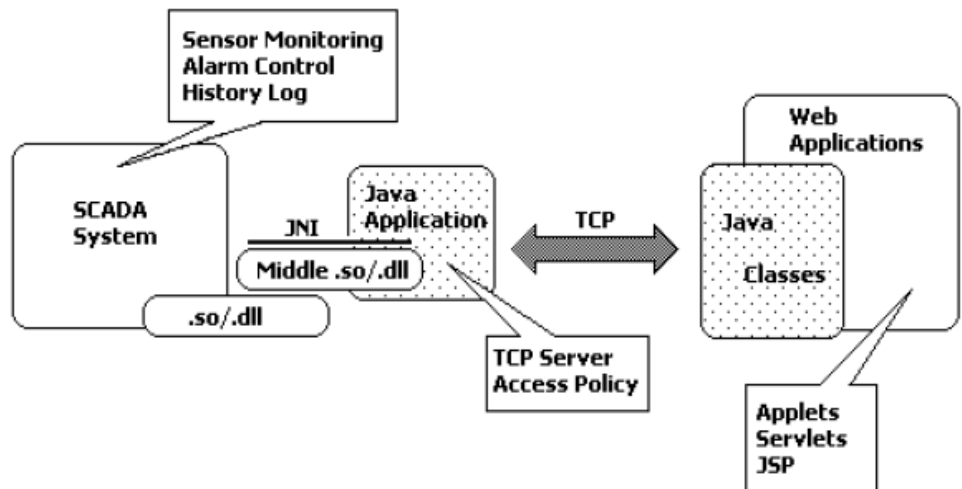

**FIGURE 8 :** Mike Clayton et al. [16] Approach

This approach shares the same disadvanteges like Duo Li et al. [3] approach, in addition to the complexity and non-flexibility in the Java application (server), because if the SCADA system changed or updated the Java application will need to be modified to use a new custom library. Also with large number of clients, the SCADA machine (Java server machine) will be very loaded with threads that may impacts the SCADA system performance. In addition, the author of this approach concentrated on the communication between the web server and the JAVA server on the SCADA machine and ignored the client to the web server (browser), how the browser updates its data with new data. There should be a data polling mechanism, which enable the browser to ask for data update.

S.H. YANG et al. [17] claimed in their guidance for web based Process control systems that Internet can be linked with the local computer system at any level in the information architecture, or even at the sensor/actuator level. These links result in a range of 4Rs (response time, resolution, reliability, and reparability). For example, if a fast response time is required a link to the control loop level should be made. If only abstracted information needed, the Internet should linked with a higher level in the information architecture such as the management level or the optimization level as shown in Figure

9. In addition, they mentioned that the Internet is a public transmission media, which is fundamentally different from other private transmission media used by many end-users for different purposes. The Internet transmission performance is associated with time delay and packet loss and possesses large temporal and spatial variation. In detail, the Internet time delay is characterized by the processing speed of nodes, the load of nodes, the connection bandwidth, the amount of data the transmission speed, etc. Therefore, it is somewhat unreasonable to model the Internet time delay for accurate prediction at every instant. In addition, they exposed to the problem of concurrent users, as they said that the special feature of the Internet-based SCADA is multiple-users and the uncertainty about users. The number and location of the users keep changing and the operators cannot see each other or may never have met. It is likely that multiple-users may try to control concurrently a particular process variable in which case some problems may arise. So, coordination among multiple-users becomes very important.

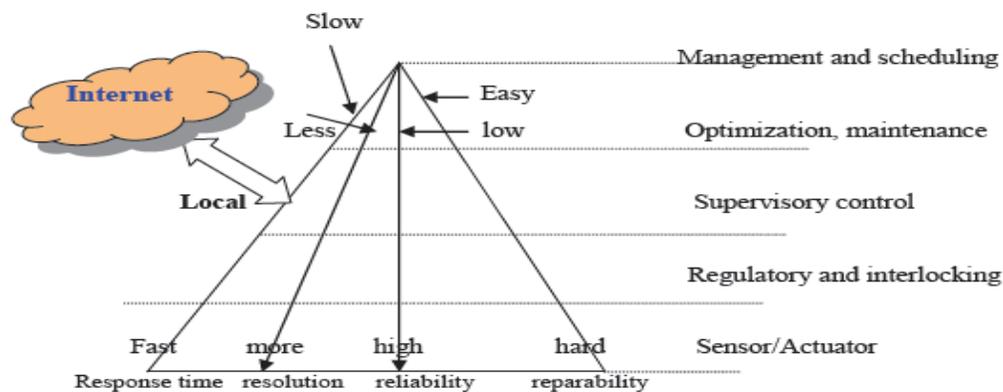

**FIGURE 9 :** S.H. YANG et al. [17] Guidance (information architecture levels)

### 2.5 AJAX

Actually, AJAX (Asynchronous JavaScript and XML) started a new line in web based SCADA systems because it enables us to create more interactive web pages, which are our way to replace traditional desktop SCADA application. AJAX considered as the future of the Internet because of its ability to simulate desktop applications by the new features it offers like asynchronous client-server calls and partial-page updates. Ajax is a group of interrelated web development techniques used on the client-side to create interactive web applications. With Ajax, web applications can retrieve data from the server asynchronously in the background without interfering with the display and behavior of the existing page. The use of Ajax techniques has led to an increase in interactive or dynamic interfaces on web pages. Truong Chau et al. [5] used AJAX in their approach to get the OPC DA data to the client web page with high user interactivity as shown in Figure 2. But because of the problems of the HTTP protocol as a stateless protocol they had to use a simple data polling mechanism by using an AJAX Timer to poll for new OPC DA data set from the OPC DA server through the web server. When using an AJAX style of programming the old, classic programming approach must be given up. There is no form submit any more that posts all the client state to the server and requests for a complete new page description using HTML. Instead of loading several pages until the functionality done, only one page loaded and stay in the browser until the end of functionality. With the Ajax model, the execution of the web page processed in three different phases: loading the page, loading data from the server using AJAX techniques and interaction with the page using AJAX techniques

The web standards used in AJAX are well defined, and supported by all major browsers. AJAX Applications are browser and platform independent. The main disadvantage of AJAX is security. Sometimes developers do not put checks on the data coming into the server - they assume that it is coming from their own website. Unfortunately, this is subject to injection attacks. Furthermore, there are several ways to "fake" how the data is coming in and making detection to all is impossible. Another disadvantage is that AJAX does not play well in encrypted environments. AJAX relies on plain text transmission (nothing but text can be transmitted through AJAX anyways), and so encrypting this stream and having the server-side program deals with it presents large problems [18].

### 3. COMPARATIVE EVALUATION

None of the mentioned approaches considered the feasibility of the realization of their work. If we look closely to [4, 8, 9, 10] we will find that they use XML and/or OPC XML-DA technologies, which have the disadvantages such as:

- Not suitable for transferring large data volumes
- XML technology is generally slower than COM
- The interaction parameters coded using XML, which leads to an overhead.
- An OPC XML-DA Service is stateless.

Actually, OPC XML-DA designed for Internet access and enterprise integration and based on its platform-independence; it mainly implemented in embedded systems and on non-Microsoft platforms. However, due to its high resource consumption and limited performance, it was not as successful as expected for this type of applications.

In [12] the authors used the features offered by Webservices to design and implement a web based SCADA system. Moreover, they could solve the challenges, which they faced such as thin client support, refresh of GUI window, and firewalls restrictions. However, they could not solve real time data collection challenge in an efficient and effective way because they used a constant refresh frequency of a few seconds, which will lead to non-synchronized process data transfer.

In [15], to allow each node (client or server) to start sending data to the other node independently at any time, the authors had to maintain a connection in a permanent open state by making the server continuously sends useful or unuseful data to the client. In addition, this will affect the performance of the system especially with large number of clients and low network bandwidth, and this may lead to server crash.

In [3] the authors will need to find a way to periodically update file2, which may be a loop or any other way. Also the java applet, which embedded in the HTML file1, will need to use a Timer periodically to get the new data from file2 and then they will face the problem of synchronization between the applet Timer interval and the process of updating file2 with new data and this will be difficult. In addition, they ignored the heavy network load and the need for large server memory to handle the large number of requests and the heavy load on the server CPU. It is very difficult (nearly impossible) to guarantee a real time behavior with such limitations. The only solution is to spend more money to increase the network bandwidth and the server CPU speed and memory capacity (The hardware resources).

In [14], the designed CyberOPC does not solve the problem of periodic data update of client to get new data. Moreover, when the remote client makes a request, it will not check if there is a change in these data from the last sent data or not so, if the control process has a high frequency data change, the remote client has to increase its periodic data requests, which affects the server efficiency and network bandwidth.

In [5], the authors did not take any attention to the efficiency of their approach. The Timer, which they used, will get the data without any care if there is a significant data change or not. In addition, how they can specify the Timer interval, there are two cases for that, first, when the interval is small i.e. one second (high frequency data polling), where the server and network will suffer. Second, if the interval is large they will loss some data change events that maybe very important and the system will not considered as a Real-Time monitoring system. Therefore, this approach still needs some modifications, which enable us to get better efficiency and real time behavior.

## 4. CONSLUSION & FUTURE WORK

None of the mentioned approaches considered the feasibility of the realization of their work. No one of them succeeded to achieve a real time behavior, which is very important in the functionality of the SCADA systems, and especially web based SCADA systems. In addition, no approach addresses the efficiency and effectiveness of the designed system. They did not pay any attention to the consumption of the system available resources such as CPU load (of the web server) and Network bandwidth. These issues are very important to get a more reliable and available SCADA system. In future, all the researchers in this area should take care of these points to make web based SCADA applications competitive to traditional desktop applications.